\documentstyle[12pt,nuclphys,psfig] {article}
\newcommand{\be}{\begin{equation}}
\newcommand{\ee}{\end{equation}}
\newcommand{\bd}{\begin{displaymath}}
\newcommand{\ed}{\end{displaymath}}
\newcommand{\ba}{\begin{array}}
\newcommand{\ea}{\end{array}}
\newcommand{\bt}{\begin{tabular}}
\newcommand{\et}{\end{tabular}}
\newcommand{\bc}{\begin{center}}
\newcommand{\ec}{\end{center}}
\newcommand{\bn}{\begin{enumerate}}
\newcommand{\en}{\end{enumerate}}
\newcommand{\bi}{\begin{itemize}}
\newcommand{\ei}{\end{itemize}}
\newcommand{\bqr}{\begin{eqnarray}}
\newcommand{\eqr}{\end{eqnarray}}
\newcommand{\bfig}{\begin{figure}[tbp]}
\newcommand{\efig}{\end{figure}}
\newcommand{\btab}{\begin{table}[ht]}
\newcommand{\etab}{\end{tabular}\ec\end{table}}
\newcommand{\bl}{\begin{large}}
\newcommand{\el}{\end{large}}
\newcommand{\vspc}{\vspace{0.3cm} \bc}

\newcommand{\hh}{\hline\hline}

\newcommand{\etal}{{\em et al.}}

\newcommand{\Ps}{P_{\sigma}}

\newcommand{\de}{\delta}

\newcommand{\vr}{\hbox{\bf r}}

\newcommand{\nuc}[2]{\mbox{\relax\ifmmode{}^{#1}{\protect\text{#2}}\else${}^{#1}$#2\fi}}

\newcommand{\citetable}[1]{table~\ref{#1}}

\title      {Rotational properties of $^{252, 253, 254}$No.\\
	     Influence of pairing correlations.}
\author     {
             T. Duguet, P. Bonche \\
             {\em Service de Physique Th\'eorique, CEA Saclay,} \\
             {\em 91191 Gif sur Yvette Cedex, France}
\and         P.-H. Heenen \\
	     {\em Service de Physique Nucl\'eaire Th\'eorique, Universit\'e Libre} \\ 
	     {\em de Bruxelles, C.P 229, B-1050 Bruxelles, Belgium}
            }

\input epsf

\begin{document}
 
\maketitle
 
\begin{abstract}

Rotational bands of $^{252, 253, 254}$No and the fission barriers of $^{254}$No at spin 0$\hbar$ and 20$\hbar$ are calculated with the Hartree-Fock-Bogolyubov theory and the Lipkin-Nogami approximate particle number projection. The SLy4 Skyrme force is used in the particle-hole channel. A zero-range force with and without density-dependence is used in the particle-particle channel. The experimental ground state deformation (${\cal{Q}}_{20}$ = 32.8 b) is reproduced as well as the increase of the dynamical moment of inertia with frequency  both for $^{252}$No and $^{254}$No. The rotational band of $^{253}$No is also calculated. Fission barriers of $^{254}$No at spin 0$\hbar$ and 20$\hbar$ show the robustness of shell-corrections  against rotation in these heavy nuclei.

\end{abstract}

\section{Introduction}
\label{secintro}

The ground-state rotational band of $^{254}$No has been observed recently~\citemany{leino,reiter}. Its properties show two interesting features. First, this nucleus has a large deformation. The deduced quadrupole deformation is $\beta_2$ = 0.27 $\pm$ 0.02, in good agreement with previous calculations~\citemany{cwiok2,patyk}. Secondly, a strong robustness of shell-effects against rotation in such heavy nuclei is expected from the calculations as the rotational band is observed up to a spin of 22$\hbar$.

Rotational properties constitute a good opportunity to test theoretical models in an extreme mass region, far away from the ones where effective forces are usually adjusted. They are sensitive to both shell effects (related to the particle-hole channel) and pairing correlations (related to the particle-particle channel), which allows to probe all the components of the force. The ability of the models and forces to describe rotational bands in a mass region just below the super-heavy elements can be a discriminating test for them to judge their predictive power in the super-heavy region.

The aim of the present study is to calculate the ground-state rotational bands of $^{252}$No and $^{254}$No which are experimentally known and of $^{253}$No, for which experiments are planned. Special attention will be paid to the influence of pairing correlations on rotational properties. A Skyrme-type interaction is used in the particle-hole channel and two parametrizations are analyzed for the particle-particle channel~: a zero-range force with and without density dependence. This latter point is particularly relevant as pairing correlations are crucial in reproducing  dynamical moment of inertia and quasi-particle Routhians.

The paper is organized as follows. Section 2 recalls the theoretical framework of the calculations and the different forces used in the particle-particle channel. In section 3, the relevance in the A $\approx$ 250 mass region of the delta density-dependent pairing force, previously fitted on the rotational bands of super-deformed nuclei in the A $\approx$ 150 mass region~\cite{rigol}, is discussed. The consistency of this fit is checked on odd-even mass differences which are more specifically related to pairing correlations. Then, we present results on the quasi-particle Routhians, single-particle energies and dynamical moment of inertia of the three Nobelium isotopes considered here. The fission barriers of $^{254}$No at spin 0$\hbar$ and 20$\hbar$ are also discussed in this section. Finally, we draw our conclusions in section~5.

\section{The theoretical framework}
\label{sectheory}

The method used to calculate rotational bands in No isotopes 
has been presented  in Ref.~\cite{tera} and applied to the study
of super-deformed rotational bands in the A $\approx$ 150~\cite{rigol} and 190 regions~\citemany{tera,fallon}.
It is based on the  self-consistent  cranked Hartree-Fock-Bogolyubov
method, with an approximate particle number
 projection by the Lipkin-Nogami prescription (HFBLN). This method is well suited to describe odd nuclei with self-consistent blocking.

In the particle-hole channel, we use a two-body force of the Skyrme-type, SLy4, which has been adjusted to reproduce also the characteristics of the infinite neutron matter and, consequently, should have good isospin properties~\cite{chab}.  This force has been shown to describe satisfactorily nuclear
properties for which it had not been adjusted such as 
 super-deformed rotational bands \cite{rigol,fallon}
and the structure and decay of
 super-heavy elements \cite{cwiok1}. It seems therefore an adequate
choice for our study of No isotopes.

In the $T$ = 1 particle-particle channel, we use two types of force for {\it n-n} and {\it p-p} pairing. First, a density {\it independent} contact interaction (eq. \ref{potvol}) fitted to reproduce the average proton and neutron pairing gaps in $^{254}$Fm \cite{cwiok2} ($V_{0, n}^{vol}$ = $-$ 250 MeV.fm$^{3}$, $V_{0, p}^{vol}$ = $-$ 290 MeV.fm$^{3}$):

\be
\hat{V_{\tau}}^{vol} =  \frac{V_{0, \tau}^{vol}}{2} \, \, (1 - \Ps) \, \, \de (\vr_1 - \vr_2 )  \, \,, \, \, 
\label{potvol} \vspace{0.3cm}
\ee
where $\Ps$ is the spin exchange operator.
In the second case, the pairing correlations have been treated with a surface-peaked delta force (eq. \ref{potsurf}) adjusted on the low spin behavior of the moments of inertia of super-deformed bands in the A $\approx$ 150 region \cite{rigol}. It is given by

\be
\hat{V_{\tau}}^{surf} =  \frac{V_{0, \tau}^{surf}}{2} \, \, (1 - \Ps) \, \, \de (\vr_1 - \vr_2 )  \, \, (1 - \frac{\rho(\vec R)}{\rho_c}) \, , \vspace{0.3cm}
\label{potsurf}
\ee
where $V_{0, n}^{surf}$ = $-$ 1250 MeV.fm$^{3}$, $V_{0, p}^{surf}$ = $-$ 1250 MeV.fm$^{3}$, $\rho(\vec R)$ is the local density of the nucleus and $\rho_c$ = 0.16 fm$^{3}$ the nuclear saturation density. In this way, this pairing interaction is mainly active in the surface region of the nucleus.

With a contact interaction, one has to define a cut-off procedure for the active pairing space; a smooth cut-off~\cite{bonche} simulates the decay of coupling as it would occur with a finite-range force. For the volume pairing, the active pairing space includes all states up to 5 MeV above the Fermi energy. For the surface pairing, the active space includes roughly one major shell, from 5 MeV above to 5 MeV below the Fermi level.

\section{A pairing exploration}
\label{secpairing}

The surface pairing interaction (see eq. \ref{potsurf}) has
 been adjusted on moments of inertia of SD bands in the A $\approx$ 150 mass 
region. Systematic calculations have shown that the same 
parametrization leads also to a very good agreement with experimental data for even, odd and odd-odd nuclei in the A = 190 region~\cite{fallon,PH}. Before using that parametrization to describe rotational
bands in No isotopes, it is worth testing its validity 
for the description of ground-state properties in the A~$\approx$~250 region.

For this purpose, we have calculated fifteen nuclei around $^{252}$No and $^{254}$No. The Nobelium isotopes from $^{250}$No to $^{256}$No, the $^{252}$No isotones from $^{250}$Fm to $^{255}$Lr and the $^{254}$No isotones from $^{252}$Fm to $^{256}$Rf. For each odd nucleus, the different 1-qp states have been self-consistently calculated and the lowest in energy has been selected. The experimental spins and theoretical spin projections on the symmetry axis (K-value) of the ground-state of these nuclei are compiled in~\citetable{tabenergie}. The calculated K-value are consistent with experimentally known (or assigned) spins. This gives some confidence in the level scheme obtained with the SLy4 parametrization and also to the spin predictions made for nuclei for which no experimental data are available. This is also supported by self-consistent calculations performed with a delta density-independent pairing  and a Woods-Saxon potential~\cite{cwiok3} which found the $^{253}$No and $^{255}$Lr spins to be K$^{\pi}$ = 9/2$^{-}$ and 7/2$^{-}$ respectively.

The next step is to choose a relevant quantity to compare theoretical and experimental pairing correlations. One can use a pure theoretical definition of the ``gap'' such as the gap at the Fermi energy $\Delta_{\epsilon_F}$, a mean gap ${\rm < \Delta >} \, = \, \frac{\sum_k \, u_k v_k \Delta_k}{\sum_k \, u_k v_k}$ or the lowest quasi-particle energy. These quantities may be compared to experimental ``equivalents'' 
defined by the excitation spectrum of even nuclei or by mass differences
between odd and even isotopes. This is a practical procedure to deal with a systematic work on pairing because it only requires the theoretical calculation of a single nucleus at a time and avoids microscopic calculations of odd nuclei when considering even ones. 

However, this procedure can be improved by imposing the following two requirements: 
\bi
\item[(1)] The quantity used to extract pairing information must be derived in the same way both from experiment and theory, \\
\item[(2)] it should be mainly sensitive to pairing correlations and eliminate as much as possible mean-field contributions to total energies.
\ei

\subsection{Odd-even mass formulas}
\label{subsecformula}

Several well known quantities which aim at evaluating the neutron or proton ``pairing gaps'' are finite-difference mass formulas \citemany{mad,jensen}. They approximate the systematic additional binding energy due to pairing of an even-even nucleus with respect to an odd-even neighbor. We give here the two-point (first order) and three-point (second order) formulas

\bqr 
\Delta_q^{(2)}(N) \, &=& \, (-1)^N \, \left[E(N-1) - E(N)\right] \, \,  \\
\label{delta2}
\Delta_q^{(3)}(N) \, &=& \, \frac{(-1)^N}{2} \, \left[E(N-1) - 2E(N) + E(N+1)\right] \, \,,
\label{delta3} \vspace{0.3cm}
\eqr
where the neighboring nuclei are taken along an isotopic (q = neutron) or isotonic chain (q = proton).

To fulfill the above requirement (1), the $\Delta_q^{(n)}$'s have to be calculated theoretically with full consistency thanks to the correct treatment of odd-N (or Z) nuclei in the model.

However, these quantities are not a measure of the pairing gaps only. They include a contribution arising from the mean-field which depends upon the order (n) of the formula and remains to be evaluated. $\Delta_q^{(2)}(N)$ can be roughly estimated as

\bqr
\Delta_q^{(2)}(N) \,  &=& \, (-1)^N \, S_q(N)   \, \\
                   &\approx& \, \Delta_q^{(2) \, pairing}(N) \, + \, (-1)^{N + 1} \, \lambda_q  \, \,  \nonumber ,
\label{delta2modif} \vspace{0.3cm}
\eqr
where $S_q(N)$ is the ``one-nucleon separation energy'', $\lambda_q$ is the neutron or proton chemical potential and $\Delta_q^{(2) \, pairing}(N)$ defines the contribution to $\Delta_q^{(2)}(N)$ coming from pairing correlations only.

 As a consequence, $\Delta_q^{(2)}(N)$ is dominated by the contribution from the mean-field since, apart for drip-line nuclei, we generally have $- \lambda \gg \Delta^{pairing}$. For $\Delta_q^{(3)}(N)$, however, the two contributions are of the same order of magnitude in light nuclei \cite{satula1} and the pairing dominates a priori in heavy ones. Actually, to isolate the influence of the pairing field, it is necessary to use high order formulas~: the higher the order of the formula, the smaller the contribution coming from the particle-hole channel.

The use of $\Delta_q^{(3)}(N)$ is sufficient as our aim is limited to test the validity of the pairing parametrizations that we use for No isotopes
without performing a global fit in the whole nuclear chart. To fit a pairing force and fulfill requirement (2), one should take a higher order formula.

\subsection{Pairing adjustment}
\label{subsecfit}


\begin{figure}[htbp]
\centerline{\epsfbox{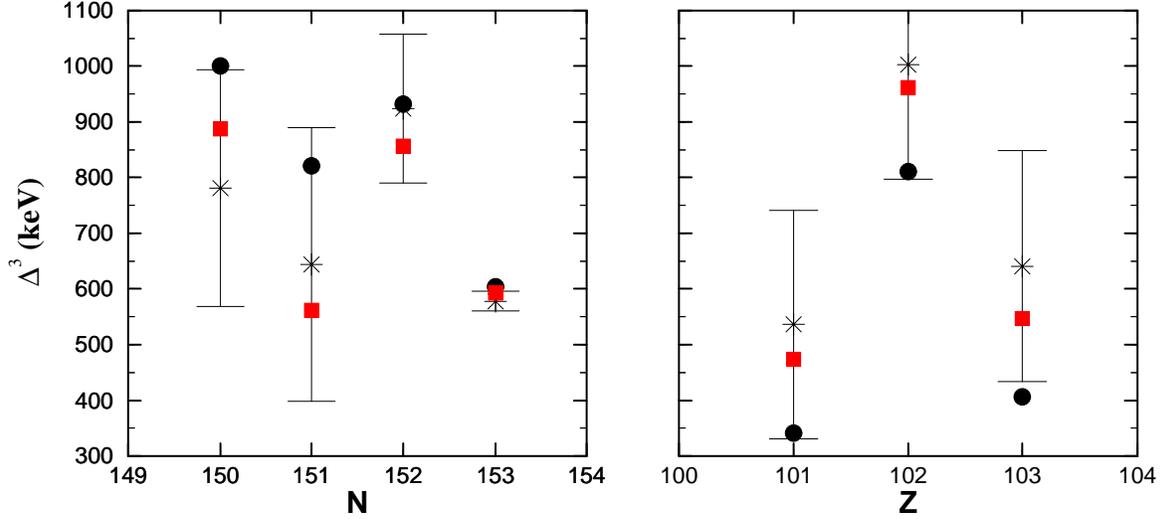}}
\caption{Three-points odd-even mass formula for three isotones (right) and four isotopes (left) of $^{254}$No. These proton and neutron ``gaps'' are given for experiment (stars) with error bars, pairing 1 (circles) and pairing 2 (squares).}
\label{figdelta}
\end{figure}

 Figure~\ref{figdelta} shows the experimental and calculated values of
 $\Delta_q^{(3)}(N)$ for neutrons along a chain of Nobelium isotopes and for 
protons along the
 neighboring isotones of $^{254}$No. The surface
 pairing fitted to reproduce correctly the SD bands in the
 A $\approx$ 150 mass region gives  satisfactory results, since all calculated
 values fall within the error bars. We will refer to this parametrization as
pairing 1.

 The experimental error bars are quite large and do not allow
a very precise adjustment of the pairing interaction. Variations
of the strengths  $V_{0,q}^{surf}$ by 100 MeV
still lead to $\Delta_q^{(3)}(N)$ within the error bars. Small
variations of the cut-off energies $E_q$ can also be explored.
 We call ``pairing 2''  the set of parameters 
 $V_{0, n}^{surf}$ = $-$
 1250 MeV.fm$^{3}$, $V_{0, p}^{surf}$ = $-$ 1350
 MeV.fm$^{3}$ and an active window of 4.4 MeV and 5.4 MeV around the Fermi energy for neutron and proton respectively. This set leads to 
theoretical $\Delta_q^{(3)}(N)$ the closest to the middle of
the experimental error bars. This second fit allows us to explore a significant range of variations for the parameters with respect to known or estimated experimental errors.

 \vspace{0.5cm}

\begin{figure}[htbp]
\centerline{\epsfbox{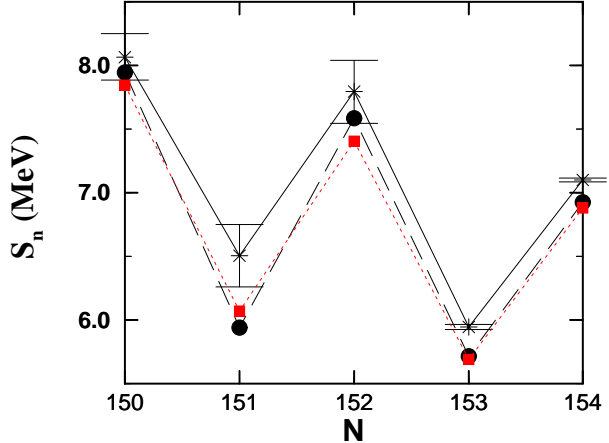}}
\caption{One neutron separation energy S$_n$ for five $^{254}$No isotopes. Stars with full line are for experiment with error bars, circles with long-dashed line for pairing 1 and squares with dotted line for pairing 2.}
\label{figsn}
\end{figure}

Figure~\ref{figsn} displays the often used one neutron separation energy $S_n(N)$ from $^{252}$No to $^{256}$No. The mean deviation from experimental data is about 300 keV. 

Surprisingly, there is no one to one visible relationship between the improvement of $\Delta_n^{pairing}$ obtained with pairing 2 (see Fig.1) and the modification that this parametrization brings on the $S_n(N)$. This can be understood by analyzing the $S_n(N)$ formula (see eq. \ref{delta2modif}). Indeed, $\Delta_n^{pairing}$ enters into $S_n(N)$ with a different sign for odd and even N. The change of $\Delta_n^{pairing}$ being the main change in  $S_n(N)$, the decrease of $\Delta_n^{pairing}$ leads to an increase of $S_n(N)$ for odd N and to a decrease for even N. In this way, the odd  $S_n(N)$ are improved contrary to the even one. Thus, there is an apparent paradox~: while improving the pairing ($\Delta_q^{(3)}(N)$), some $S_n(N)$ are spoiled because of this alternating sign. This result illustrates the interplay between the pairing and mean-field parts of the energy. The mean field is responsible for the global shift of the $S_n$ with respect to experiment. Thus, if it were corrected, the decrease of $\Delta_n^{pairing}$ would also improve the even $S_n$, since they would be too high with pairing 1 in this case. 

Consequently, the use of $S_n(N)$ to evaluate the influence of the pairing adjustment can reveal a weakness of the mean-field since this quantity is sensitive to both particle-particle and particle-hole channels (second order formula). \vspace{0.5cm}

\begin{figure}[htbp]
\centerline{\epsfbox{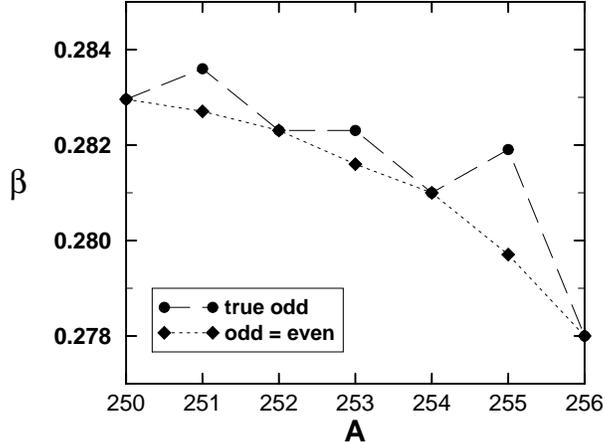}}
\caption{Quadrupole deformation $\beta_2$ for ground-state isotopes of $^{254}$No calculated with pairing 1. Circles with long-dashed line are for the full self-consistent treatment of odd nuclei. Diamonds with dotted line are for odd nuclei calculated self-consistently without blocking.}
\label{figbeta}
\end{figure}

Figure~\ref{figbeta} gives the quadrupole deformation $\beta_2 = \sqrt{\frac{\pi}{5}} \frac{< \hat{Q} >}{A < r^2 >}$ along the isotopic chain using pairing 1. All these nuclei have
axial quadrupole deformations in their ground state. The values obtained with pairing 2 differ by at most 0.5~$\%$ for all isotopes
although  the neutron gap are modified by up to~30~$\%$.
The value obtained for
 $^{254}$No is close to the one deduced from 
experiment~\citemany{leino,reiter}
 ($\beta$ = 0.27 $\pm$ 0.02). This result is supported by the ones obtained in  HFB calculations 
with SLy4 and volume pairing~\cite{cwiok2}, with the
 Gogny force~\cite{egido}, by the
relativistic Hartree-Bogoliubov method~\cite{lala} and by a
macroscopic-microscopic
 approach~\cite{muntian}.  Taking the corrections due to $\beta_4$ into account according to Ref~\cite{hass}, one obtains $\beta_2$ = 0.264. 

The deformation along the whole line between $^{250}$No and $^{256}$No shows a very smooth global decrease with increasing $A$ toward shell closure at N = 184~\cite{cwiok2}. Besides this global trend the quadrupole deformation presents an odd-even staggering along the isotopic line. This very weak effect shows that even nuclei are comparatively less deformed than their odd neighbors. Figure~\ref{figbeta} also gives the quadrupole deformation for the same isotopes except that the odd nuclei are calculated self-consistently as if they were even ones without any blocking, i.e. without breaking time invariance. In this case, no staggering occurs and only the smooth decrease can be seen. The additional deformation of an odd nucleus with respect to its virtual even partner is thus directly connected to the blocking which was, up to now, known to be responsible for the odd-even staggering of the binding energy. One can thus assess that the blocking procedure, through the weakening of the pairing correlations and the polarization of the core, is responsible for a second staggering effect on the deformation.

\section{Rotational properties}
\label{secrot}

\subsection{Routhians}
\label{routh}

 \vspace{0.5cm}

\begin{figure}[htbp]
\centerline{\epsfbox{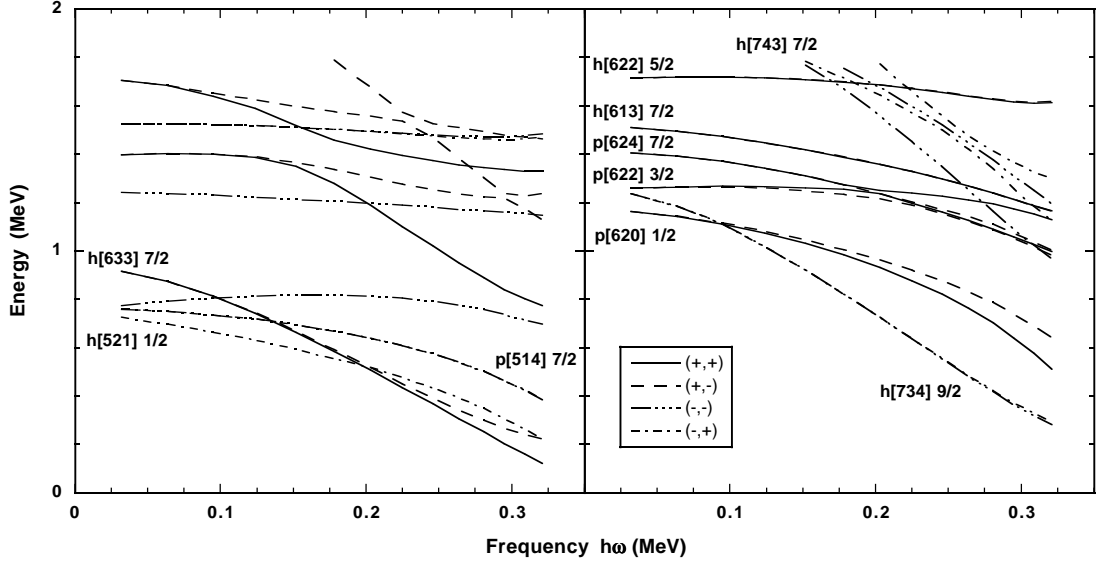}}
\caption{Quasi-particle Routhians for $^{254}$No. Proton quasi-particles are
 shown on the left panel and neutron qp on the right one. The
 conventions for parity and signature are given on the figure.}
\label{figqp}
\end{figure}

\begin{figure}[htbp]
\centerline{\epsfbox{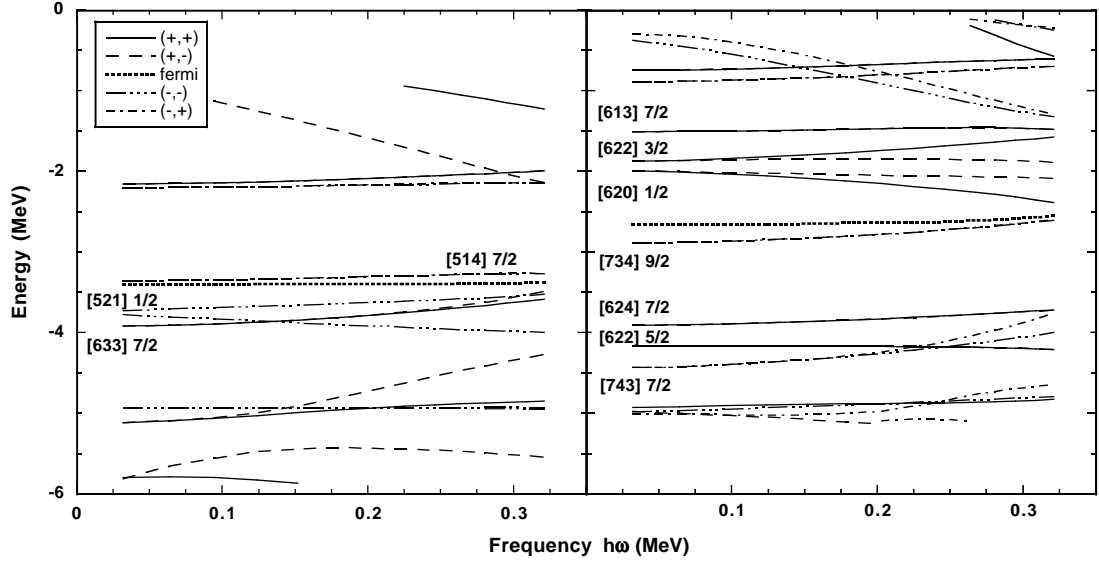}}
\caption{Single-particle spectra for $^{254}$No; protons are shown on the
 left and neutrons on the right.  The conventions for parity and signature are given on the figure. The states are
 labeled by  their dominant Nilsson component.}
\label{figsp}
\end{figure}

Figure~\ref{figqp} shows the proton and neutron qp
 Routhians of $^{254}$No and Figure~\ref{figsp}
the single-particle energies in the HF basis 
 as a function of the frequency. The
 quasi-particles (qp) are labeled by  their dominant Nilsson component
 in the HF basis and by the letter p (particle) or h (hole) indicating
 whether this component lies above or below the Fermi level.  

The quasi-proton [633]7/2$^+$ orbitals originate from the intruder
orbitals $1i_{13/2}$. As expected, they are strongly down-slopping
and present a signature splitting of 0.2 MeV for $\hbar \omega$=0.3 MeV. 
They should play an important role at high spins
in the heavy isotones of $^{254}$No, while the
 [514]7/2$^-$ qp should dominate in the light
isotones. 

The neutron quasi-particle orbitals corresponding to the [734]9/2$^-$
intruder states have the lowest excitation energies for hole type orbitals. The ground state of $^{253}$No is based on this
qp excitation. The other low lying qp
excitations are of particle type. Consequently, several 1qp excitations should be
close in energy in  $^{255}$No, leading to a large fragmentation
of the population of the rotational bands. \vspace{0.5cm}

 Large proton and neutron gaps can be seen on  Figure~\ref{figsp}  at 
all frequencies. For the neutrons, they correspond to the lighter
isotopes with N = 150 and 152 and for the protons to the heavier
isotones  with Z = 104 and Z = 108 although in this last case, the
gap disappears above $\hbar \omega$ = 0.2 MeV. 
We also obtain a gap for N~=~162, which confirms the existence of such
a deformed shell closure found in previous microscopic~\cite{cwiok2} and macro-microscopic calculations~\cite{patyk}.

\subsection{Moments of inertia}
\label{inertia}

\begin{figure}[htbp]
\centerline{\epsfbox{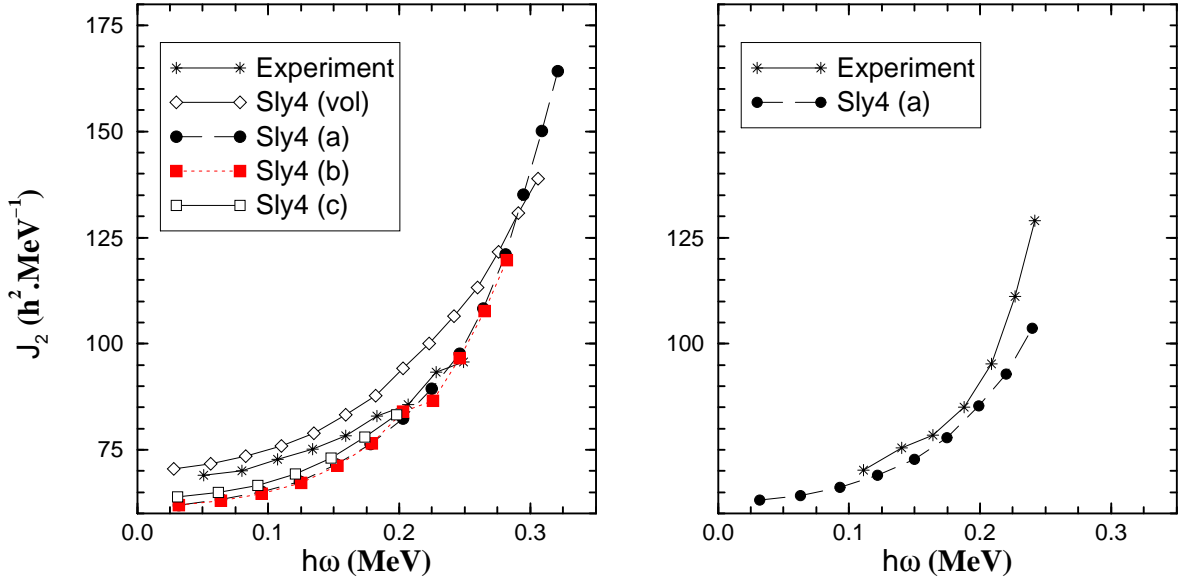}}
\caption{Dynamical moment of inertia ${\cal J}^{(2)}$ for $^{254}$No (left) and $^{252}$No
 (right). (vol) stands for the volume
 pairing, (a) for pairing 1, (b) for pairing 2 and (c) is a fit which
 differs from pairing 1 only by the value of $V_{0, n}^{surf}$ (see text).}
\label{figj2}
\end{figure}

We compare in Figure~\ref{figj2} the theoretical
dynamical moment of inertia of
 $^{252}$No and $^{254}$No with the experimental data~\cite{reiter}.
For both nuclei, we have used the pairing 1 presented in
section \ref{subsecfit}. 
In addition, we have tested three other pairing
parametrizations for $^{254}$No: the volume pairing 
defined in eq. \ref{potvol}), the  pairing 2 
discussed in section \ref{subsecfit}
and a third surface pairing in which the neutron and proton strengths
have been arbitrarily decreased to  $V_{0, n}^{surf}$= $-$ 1200 MeV.fm$^{3}$. 
This last set has been introduced to determine whether the agreement
with the data of the set 1 could be improved at low frequency staying within the error bars for the
ground states.

The agreement with the data is quite good in all cases.
The volume pairing  slightly overestimates
the moment of inertia of $^{254}$No while all the surface pairing
parametrizations slightly underestimate it at low frequencies. At higher frequencies, different behaviors for the two pairings are seen with a faster increase of ${\cal J}^{(2)}$ associated to the surface pairing.
Reasonable changes of the pairing strength do not modify significantly
the moments of inertia which seem to be more sensitive
to the region of the space (surface or volume) where the pairing is
active than to variations of its strength. An intermediate behavior
between the extreme surface and volume pairings might improve
the agreement with the data. It does not seem however appropriate to
perform such a detailed fit of the pairing interaction by looking to
properties only in this mass region. \vspace{0.5cm}

\begin{figure}[htbp]
\centerline{\epsfbox{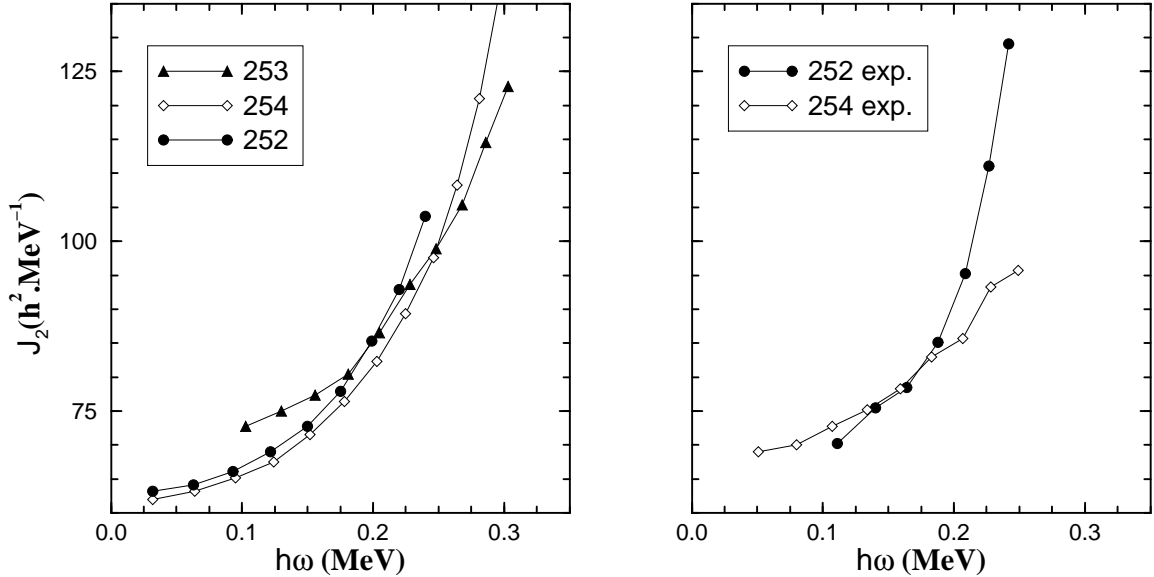}}
\caption{Dynamical moment of inertia ${\cal J}^{(2)}$ for $^{252}$No, $^{253}$No
 and $^{254}$No calculated with pairing 1 on the left panel; experimental dynamical of inertia for  $^{252}$No and  $^{254}$No are on the right panel.}
\label{figj2253}
\end{figure}

There is no experimental data
for the dynamical moment of inertia of $^{253}$No. 
The predicted ${\cal J}^{(2)}$ for $^{252}$No, $^{253}$No and $^{254}$No
 using pairing 1 are shown on Fig.~\ref{figj2253}.
The calculated rotational band of $^{253}$No is based on the intruder qp [734]9/2$^-$.
One observes an increase of the moment of inertia of $^{253}$No 
compared to $^{252}$No and $^{254}$No at low frequency, due to the
 reduction of the pairing by blocking.
 At high frequency, the high $j$
 blocked state still contributes strongly to the alignment but without
 contributing to the loss of pairing energy.
 This effect makes the dynamical moment of inertia of $^{253}$No flatter than those of $^{252}$No and $^{254}$No at high spin. 

The comparison between the moment of inertia of the two even nuclei is also fruitful. At low frequency, the two moments of inertia are identical, but ${\cal J}^{(2)}(^{252}No)$ increases faster than ${\cal J}^{(2)}(^{254}No)$ beyond 200 Kev. This feature is observed on Fig.~\ref{figj2253}. It is however more pronounced in the experimental data (see right hand side of Fig.~\ref{figj2253}). The present calculations are thus able to reproduce qualitatively this peculiar feature of Nobelium rotational properties, but more experimental data at higher spin would be of interest for quantitative comparisons.

\subsection{Fission barrier}
\label{barrier}

  The rotational band of $^{254}$No
 has been observed up to spin 22$\hbar$ and an excitation energy larger than 6 MeV at that spin~\cite{reiter2}. To understand the stability of this band, we have calculated the fission barrier of $^{254}$No
 and its dependence on angular momentum.

\begin{figure}[htbp]
\centerline{\epsfbox{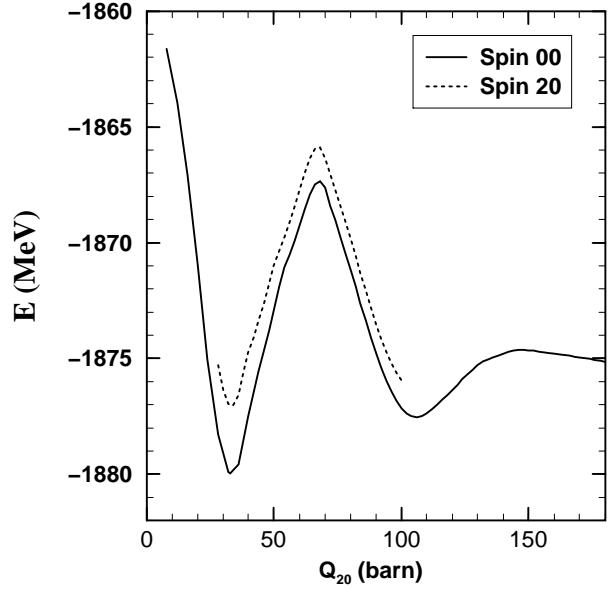}}
\caption{Fission barriers of $^{254}$No for the ground state (full line) and at spin 20$\hbar$ (dotted line).}
\label{figbar}
\end{figure}

The result is shown on Figure~\ref{figbar} for a spin of 0$\hbar$ and at spin $20\hbar$. The calculation is performed with the same formalism used for the rotational band with an additional constraint on the quadrupole moment. 

At spin 0$\hbar$ two minima are observed; a first very deep one at ${\cal{Q}}_{20}$ = 32.8 b and a second one, less pronounced, at ${\cal{Q}}_{20}$ = 106 b with an excitation energy of 2.4 MeV. These two minima are separated by a barrier at ${\cal{Q}}_{20}$ = 68 b whose energy is 12.6 MeV above the first minimum when constraining to axial deformation only. Allowing for triaxial deformations, the energy curve is only modified in the region of the first barrier which is lowered by 3 MeV. At the maximum of the barrier, the triaxial angle $\gamma$ reaches 10 $^{\circ}$. Apart for the first barrier, the nucleus is axially symmetric at all deformations. 

The height of the second barrier is 2.9 MeV with respect to the second minimum. Consequently, this second minimum could hold a super-deformed state stable against fission. However, the inclusion of octupole deformation would strongly decrease its fission half-life~\citemany{berger,egido}. Moreover, this state is very likely unstable against neutron emission. \vspace{0.5cm}

At $20\hbar$, the structure of the barrier is similar, although the two minima are slightly closer in energy and the height of the barrier is decreased to 11.1 MeV and should be lowered by a few MeV by the inclusion of triaxiality. These calculations thus confirm what has been found experimentally, namely a barrier height $\geq$ 5 MeV for spin I = $20\hbar$. This structure should remains the same up to spin $30-40\hbar$. A similar result result has been found with the Gogny force~\cite{egido}.

\begin{figure}
\begin{center}
\leavevmode
\centerline{\psfig{figure=tot32a.epsi,height=4.8cm} \psfig{figure=tot68a.epsi,height=4.8cm} \psfig{figure=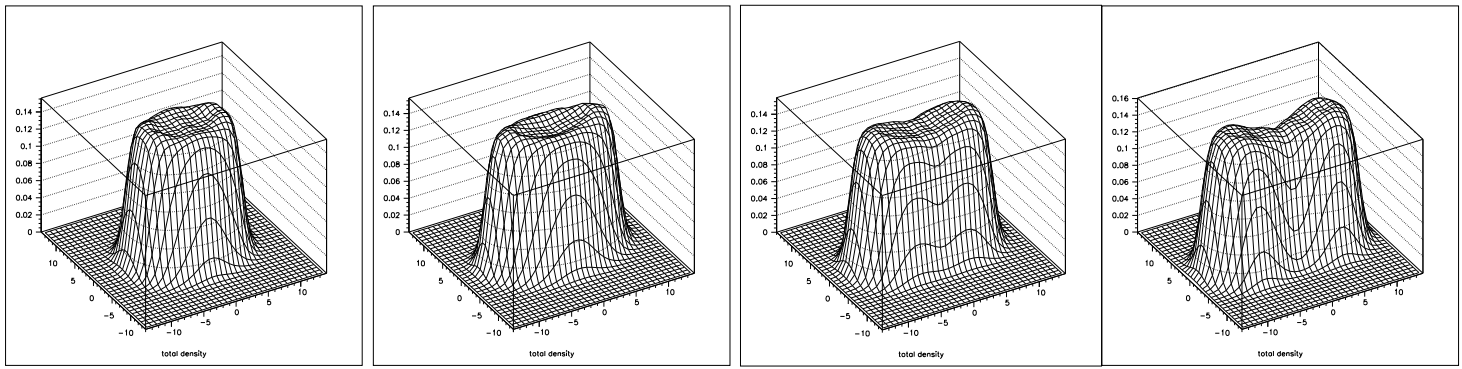,height=4.8cm}\psfig{figure=tot180a.epsi,height=4.8cm}}
\end{center}
\caption{$^{254}$No total density in y-z plan (z is the ``deformation'' axis) for four quadrupole deformations along the fission path. From the left to the right~: ${\cal{Q}}_{20}$~=~32, 68, 106, 180 barn. The values on the y-z axis are given in Fermi.}
\label{densite1}
\end{figure}

Figure~\ref{densite1} gives the $^{254}$No ground-state density distribution for four different quadrupole axial deformations. These densities are given in a plane containing the z symmetry axis. A hollow in the center of the nucleus can be seen for ${\cal{Q}}_{20}$ = 68 b (maximum of the barrier) and becomes deeper with deformation. At larger quadrupole deformations a strong increase of negative axial hexadecapole deformation takes place (see the right panels of Fig.~\ref{densite1}). Beyond the second minimum, namely for values of ${\cal{Q}}_{20}$ illustrated by the rightmost panel of Fig.~\ref{densite1}, important octupole deformations are expected.

\begin{figure}
\begin{center}
\leavevmode
\centerline{\psfig{figure=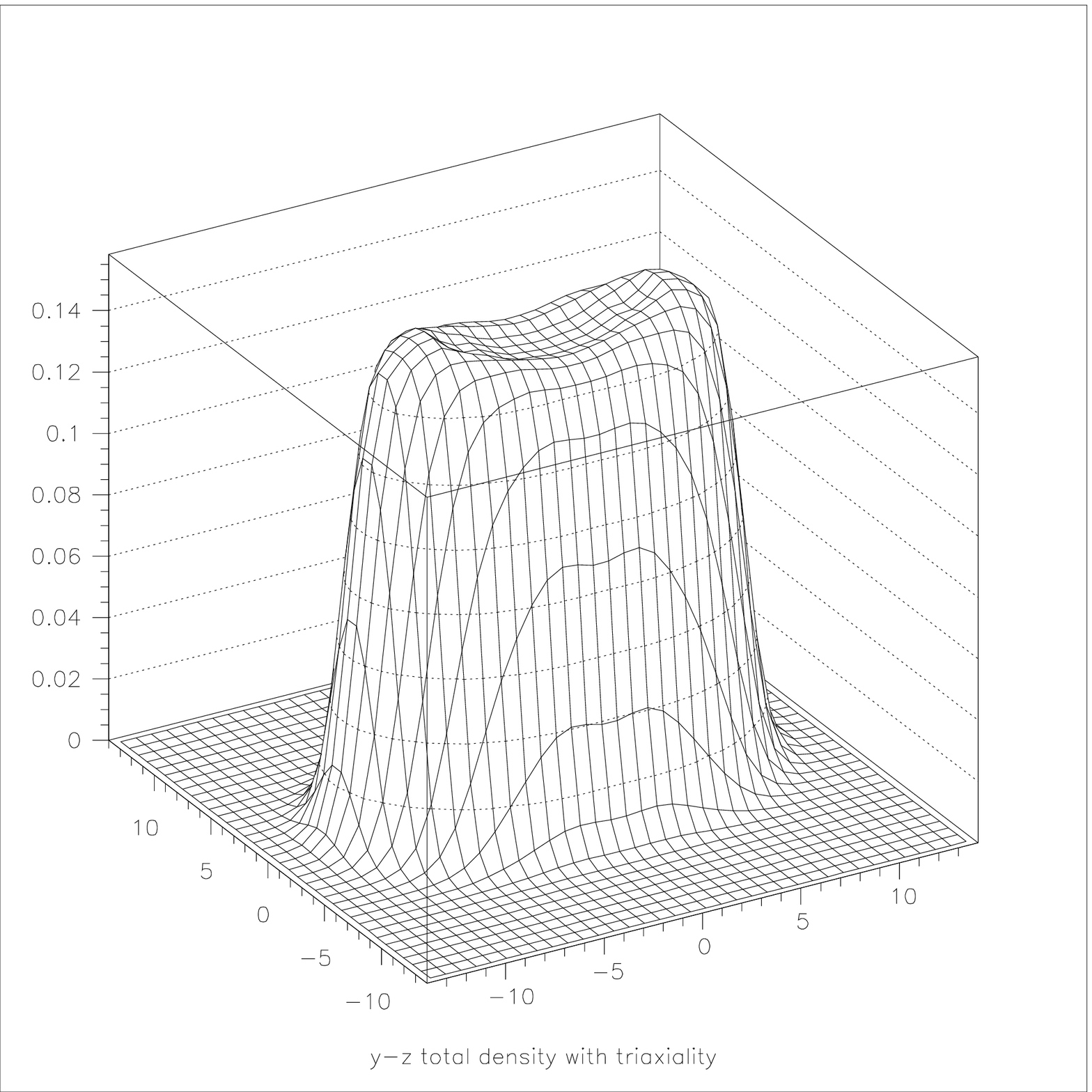,height=5cm} \psfig{figure=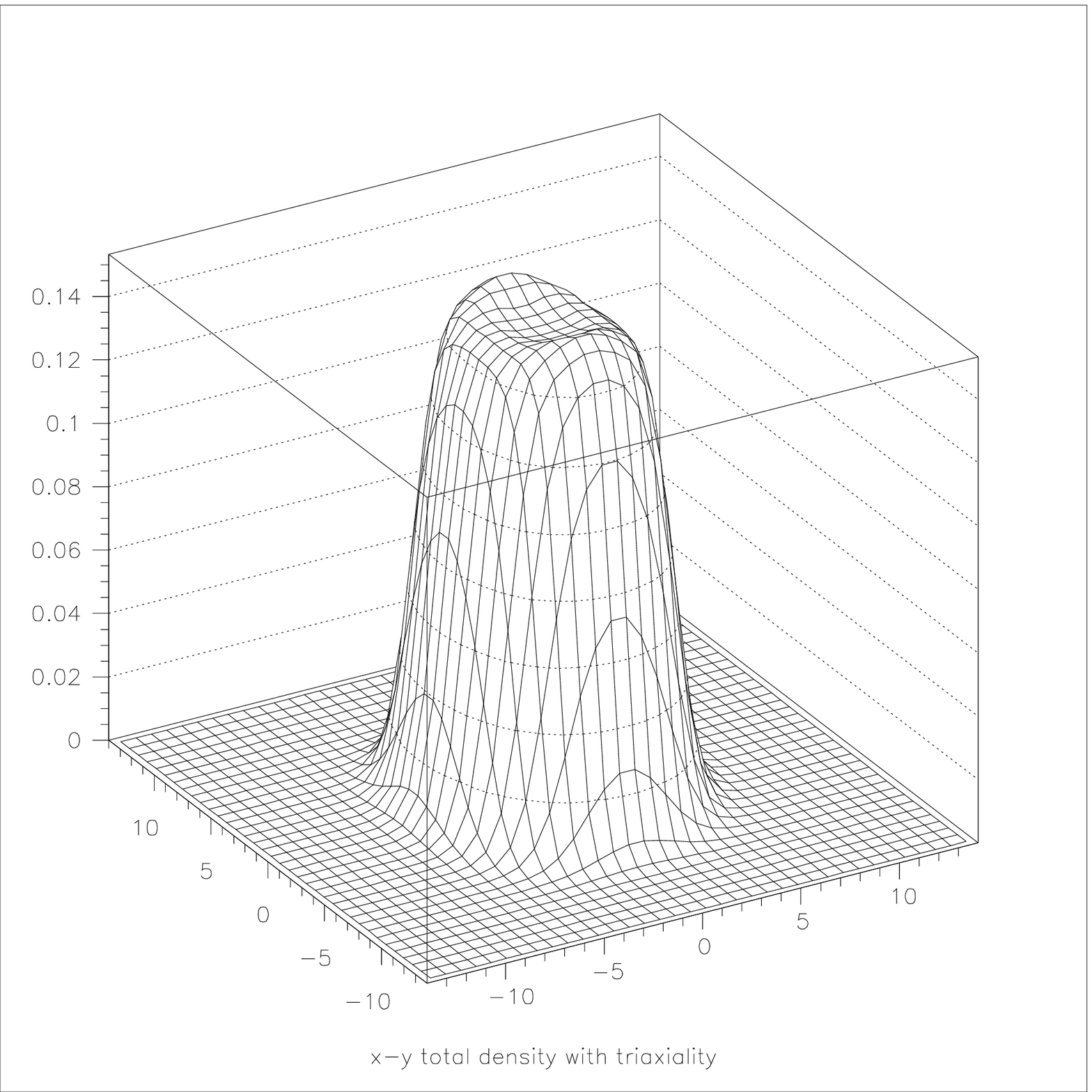,height=5cm}}
\end{center}
\caption{$^{254}$No total density in y-z and x-y plans (z is the ``deformation axis'') for ${\cal{Q}}_{20}$~=~68 b. These densities correspond to $\gamma$ $\approx$ 10 $^{\circ}$. The values on the y-z and x-y axis are given in Fermi.}
\label{densitetr}
\end{figure}

As mentioned above, triaxial deformations are important at the barrier. Figure~\ref{densitetr} shows the density distribution for ${\cal{Q}}_{20}$~=~68 b in the same plane (left panel) and in the transverse plane (right panel) for a triaxial calculation ($\gamma$ $\approx$ 10 $^{\circ}$) corresponding to the saddle point between the two minima. \vspace{0.5cm}

\vspace{0.5cm}

\section{Conclusion}
\label{secconclu}

In this paper, we have calculated rotational properties of $^{252, 253, 254}$No using the HFBLN formalism in the coordinate representation~\cite{tera}. 

In the particle-particle channel, we have used a zero-range density-dependent pairing force adjusted on super-deformed bands in the A $\approx$ 150 mass region~\cite{rigol}. We have checked the validity of this force in the present region of Nobelium using odd-even mass formulas. This previous fit gives results in agreement with the data, within the errors bars. The spins of even-odd nuclei obtained for nuclei around the Nobeliums are in good agreement with experiment. This gives credit to the single-particle spectra obtained with the SLy4 Skyrme parametrization in this mass region. Compiling ground-state quadrupole deformation for the Nobeliums isotopes between $^{250}$No and $^{256}$No, we have found a theoretically interesting odd-even staggering which originates, as the odd-even staggering of the binding energy, in the weakening of the pairing correlations and polarization of the core due to the blocking effect in odd nuclei.

We have calculated the dynamical moments of inertia corresponding to the two experimentally known rotational bands of $^{252}$No and $^{254}$No. We have also calculated the $^{253}$No's one for which experiments are planned. We have found a good agreement with experiment for two kinds of pairing force, namely the zero-range force and the zero-range force with a density dependence (active at the surface of the nucleus). We have used several fits for the zero-range density-dependent pairing force, however the dynamical moments of inertia revealed to be less sensitive to the fit used than to the region in space where the pairing is active. The above results are of particular interest because single-particle energies and pairing correlations are deeply involved in the description of rotational properties. Consequently, this gives a strong credit to the forces and fits used in the two channels, in a mass region just below the super-heavy elements.

Finally, we have calculated the fission barriers of $^{254}$No at 0$\hbar$ and 20$\hbar$ to understand the observation of its rotational band up to spin 22$\hbar$ and energy excitation around 6 MeV~\cite{reiter2}. The height of the first barrier is about 12.5 MeV at 0$\hbar$, and still around 11 Mev at spin 20$\hbar$. We expect it to remain important enough to hold a stable state against fission up to spin 30-40$\hbar$. 

Thus, in the language of macroscopic-microscopic models~\cite{strut}, we can assess that the shell-correction effects are sufficiently robust against rotation to explain the observation of the rotational band of $^{254}$No up to such high spins. We have also calculated the densities for several quadrupole deformations of $^{254}$No. These densities show a hollow in the center of the nucleus already for the ground-state and which increases with deformation. Such a hollow may affect the Coulomb part of the liquid drop energy in macroscopic-microscopic models.

\section{Acknowledgment}
\label{secremer}

We would like to thank J.L. Egido for fruitful discussions and T.L. Khoo for providing us results before publication.

\newpage


\newpage

\vspace{3mm}
\begin{large}
\centerline{\bf Tables}
\end{large}
\vspace{1cm}

\btab
\caption[T2]{Ground state experimental binding energies (in MeV) and spins of the Nobelium isotopes and isotones around $^{252}$No and $^{254}$No. Uncertainties on experimental binding energies, and theoretical spin projections on the symmetry axis are also given.
$^*$ extrapolated value~\cite{audi}.
$^{**}$ no experimental data available.
$^{***}$ experimental date taken from~\cite{audi2} .
Experimental energies and spins are taken from G. Audi \etal~\cite{audi}.}
\label{tabenergie}
\vspc
\begin{tabular}{rrrrrrrrr}
\hh
        &   &   Exp & &  & Theory   &  & \\
\hline
	&   & J$^{\pi}$ & &  &  K$^{\pi}$ & &                 \\
\hline
$^{250}$No &  &  0$^+$           & & & 0$^+$    &  \\
$^{251}$No &  &  $^{**}$         & & & 7/2$^+$  &  \\
$^{252}$No &  &  0$^+$           & & & 0$^+$    &  \\
$^{253}$No &  &  9/2$^-$$^{***}$ & & & 9/2$^-$  & \\
$^{254}$No &  &  0$^+$           & & & 0$^+$    &  \\
$^{255}$No &  &  1/2$^+$$^{***}$ & & & 1/2$^+$  & \\
$^{256}$No &  &  0$^+$           & & & 0$^+$    & \\
$^{250}$Fm &  &  0$^+$           & & & 0$^+$    & \\
$^{252}$Fm &  &  0$^+$           & & & 0$^+$    & \\
$^{251}$Md &  &  $^{**}$         & & & 1/2$^-$  & \\
$^{253}$Md &  &  $^{**}$         & & & 1/2$^-$  & \\
$^{253}$Lr &  &  $^{**}$         & & & 7/2$^-$  & \\
$^{255}$Lr &  &  $^{**}$         & & & 7/2$^-$  & \\
$^{254}$Rf &  &  0$^+$           & & & 0$^+$    & \\
$^{256}$Rf &  &  0$^+$           & & & 0$^+$    & \\
\hh   				
\etab

\end{document}